\documentclass[10pt,twoside]{article}

\usepackage[left=3cm,right=3cm]{geometry}
\usepackage{graphicx}

\usepackage{amssymb}
\usepackage{amsthm}

\usepackage{pgfplots}
\usepackage{float}
\newlength\fheight
\newlength\fwidth
\usepackage{amsmath}
\usepackage{physics}

\usepackage{bm}

\usepackage{subcaption}
\usepackage{mwe}

\usepackage{varioref}
\usepackage{hyperref}
\usepackage{cleveref}
 \usepackage{authblk}

\title{A simplified static frequency converter model for electromechanical transient stability studies of $16\frac{2}{3}$~Hz railways}
\author[1]{John Laury}
\author[1]{Lars Abrahamsson}
\author[1]{Math H.J. Bollen}
\affil[1]{Lule{\aa} University of Technology, Electric Power Engineering Group, Skellefte{\aa}}
\date{}                     
\begin{document}
\maketitle
	\begin{abstract}
		With increased share of Static Frequency Converters (SFCs) in 16$\frac{2}{3}$~Hz railway grids concerns about stability have increased. Stability studies for such low-frequency railway grids are few, and models that describe SFC dynamics are especially few.
		
		This paper presents an open SFC model for electromechanical stability studies in the phasor domain, suited for 16$\frac{2}{3}$~Hz synchronous railway grids. The developed and proposed SFC model is implemented in MatLab Simulink, together with grid and loads. Numerical studies are made, in which the proposed SFC model is validated against both measured RMS-phasor amplitude of  voltage and current at the railway grid side of an SFC. The SFC model developed is able to reproduce the measured RMS voltage and current with an acceptable accuracy.	
	\end{abstract}
	
%
	




\section{Introduction} \label{Sec: Introduction}

Low-frequency railway grids exist in Austria, Germany, Norway, Sweden, Switzerland and in some parts of the North Eastern U.S. \cite{Steimel2012}. Because the frequency in the railway grid differs from the one in the public power grid, conversion of frequency is needed. This conversion can either be done by using a Rotary Frequency Converter (RFC)  or a Static Frequency Converter (SFC) \cite{Lars2012,steimel2008,tomp1898,Pfeiffer1997}. The RFC is essentially a three-phase motor and a single phase generator mounted on the same mechanical shaft. The motor is either an asynchronous motor or a synchronous motor, depending on if the railway grid is synchronous to the public grid or asynchronous to the public grid. Therefore, two different types of low-frequency grids can be distinguished. 

If the motor of the RFC is of synchronous type, the railway grid is synchronously coupled with the public grid. This type of RFC are called synchronous-synchronous RFC. The pole ratio between the synchronous motor and synchronous generator is 3 (Norway, Sweden) or 2.4 (U.S.). Therefore, the induced frequency in the single-phase generator stator will be 16$\frac{2}{3}$ Hz in Norway and Sweden, and 25~Hz in the U.S. 

Using a three-phase induction motor on an RFC results that the railway grid is asynchronous to the public grid. This type of RFC are called asynchronous-synchronous RFC. The pole ratio between the induction motor and single-phase generator is three. The induction motor is a doubly-fed one and the slip is controlled, which results that the frequency of railway grid is controlled.

As the railway grid of Norway and Sweden are synchronously to the public grid, the frequency on the railway grid is determined by the frequency in the public grid (the ratio being exactly three). Therefore, the active power supplied from the three-phase grid to the railway grid through a synchronous-synchronous RFC is dependent on the phase angle difference between the three-phase public grid and the railway grid at the point of connections \cite{Abrahamsson117,Olofsson1995}. Only the voltage magnitudes on the three-phase grid and the railway grid sides are controlled, alternatively the reactive power generated on each side.


Since the introduction of power electronics in the 1950s \cite{Ostlund2012}, a large part of the added frequency conversion units in the Swedish railway grid is made up by SFCs. RFCs are still in use to a large extent, and some converter stations have both SFCs and RFCs in parallel operation.

The most common SFC type in the Swedish low-frequency railway is a three-phase rectifier and a single-phase inverter with a common DC link capacitance between. In the Swedish synchronous railway, the SFC inverter is controlled to mimic the steady state behaviour of an RFC. The SFC rectifier is controlled to keep constant DC voltage and control the AC voltage of three-phase feeding grid, alternatively control reactive power.

For the few electromechanical-stability studies done on synchronous railway grids, very few models exist that describe the dynamical behaviour of an SFC. One SFC model is presented in \cite{Eitzmann1997} for an electromechanical transient stability study of part of the U.S. synchronous railway grid. Based on the electromechanical time constant of an RFC, an SFC with both phase angle and voltage control is presented. However, no data is given about the parameters used. Furthermore, the control is based on a cylindrical type of synchronous generator, which is not very commonly used in synchronous low-frequency railways.

With an increased share of SFCs in the Swedish railway grid, concerns about stability have increased. The main concerns are about the decreased ratio between installed power and inertia. 
However, there are very few SFC models for synchronous railway grids and most of them belong to commercial softwares. Therefore there is a need for an open SFC model that can be used for electromechanical stability studies for synchronous railway grids.


The modelling approach used in this paper has been to adapt three-phase VSC-HVDC models, of which there is a large amount of in the literature. As the SFC model developed is intended for electromechanical stability studies in the phasor domain, VSC-HVDC models capturing such dynamics have been adapted from references such as \cite{HVDCbook2016,JingLiu2009,Machowski2008}. VSC-HVDC models for electromechanical stability studies in phasor domain have several benefits, for example Phase Lock Loop (PLL) can be omitted or simplified \cite{Liu2014UseThis,paulo2012,Shewarega2014}. Furthermore, the computational burden of such models is smaller compared to the time-domain models that are needed for Electro Magnetic Transient (EMT) simulations.



The aim of this paper is to present a simplified SFC model with current limitation control for electromechanical transient stability studies in the phasor domain. The SFC model is validated against both RMS voltage and current from measurements of an inverter of an SFC operating in the Swedish railway grid.  The proposed SFC model provides an adequate description of the railway side dynamics of the SFC as it is able to reproduce the measured data. 

The remainder of this paper is organised as follows: In \Cref{Sec: SteadyState} the SFC model for electromechanical stability studies is presented. \Cref{Sec: Implemenation} describes how the SFC is implemented for simulation in a grid. The data used is also presented in the same section. 
The validation of the model is done in \Cref{Sec: Validation} and \Cref{Sec: Conclusions} concludes the paper.

\section{SFC modelling} \label{Sec: SteadyState}
\begin{figure}
\centering
\includegraphics[width=1\textwidth,keepaspectratio]{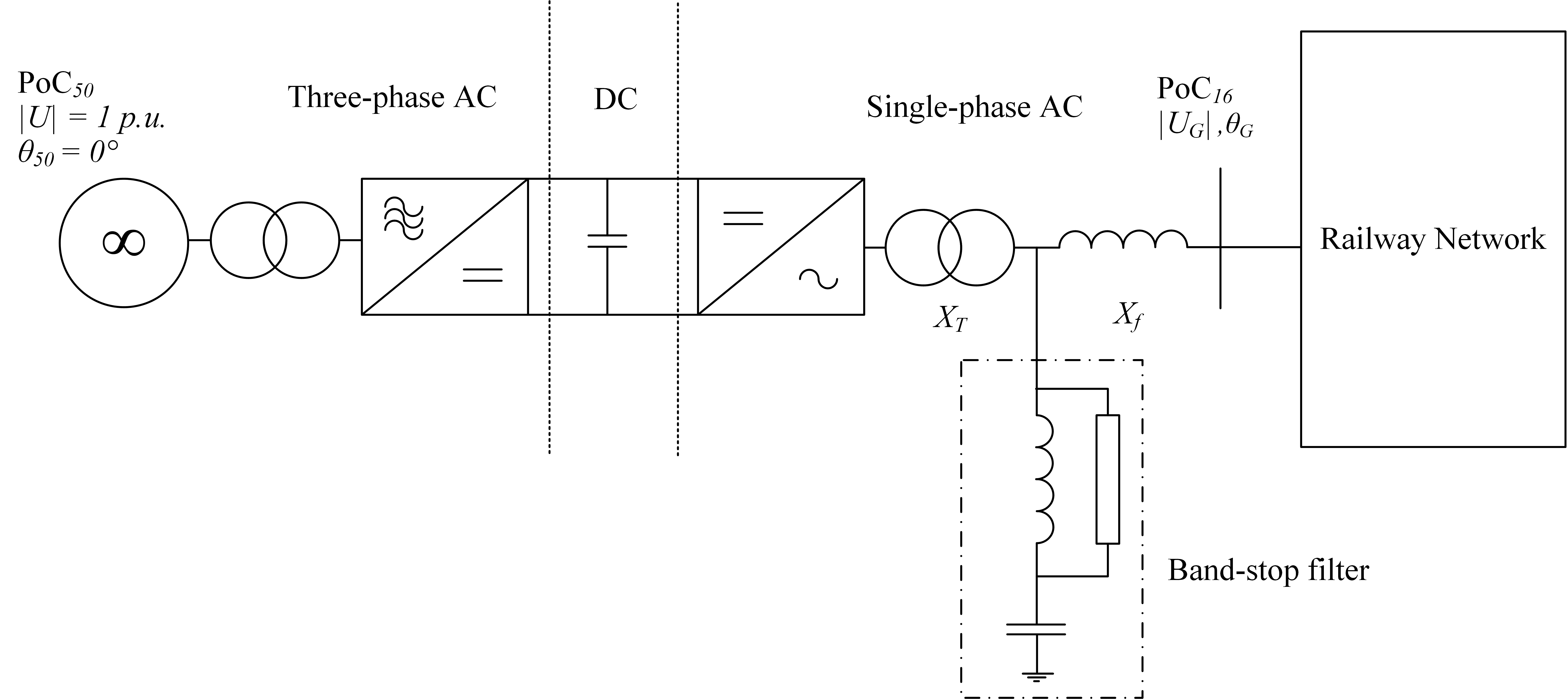} 
\caption{General schematic of an SFC connected to both three-phase grid and single-phase grid.}\label{Fig: SFC_circuit}
\end{figure}

The control objective of the SFC at the railway side is to regulate its voltage magnitude and phase such that the steady-state behaviour of an RFC is obtained at the Point of Connection at the railway grid side, denoted $\text{PoC}_{16}$, see \Cref{Fig: SFC_circuit}.
The SFC is connected to the railway grid through a transformer reactance $X_T$, an inductive filter $X_f$ and a band-stop filter as seen in \Cref{Fig: SFC_circuit}. 

\subsection{Simplifications and modelling approximations}
For simplicity, it is assumed that the DC capacitance in \Cref{Fig: SFC_circuit} is large enough to always keep constant DC voltage, and rectifier controls can then be neglected.

The SFC is on the three-phase side connected to an infinite bus, where both the 50~Hz voltage level and phase angle is constant. This is justified as most RFCs and SFCs are often connected to strong parts of the three-phase public grid. It simplifies the modelling and simulation of the three-phase grid side. 

The voltage phase angles of the three-phase grid and single-phase grid are measured continuously via a PLL for respective grid side. However, as the simulations are done in phasor domain, the phase angles can be obtained directly from the simulations. The PLL is therefore assumed ideal in the modelling, in other words no time-delay is introduced by the PLL.

\begin{figure}
\centering
\includegraphics[scale=0.9]{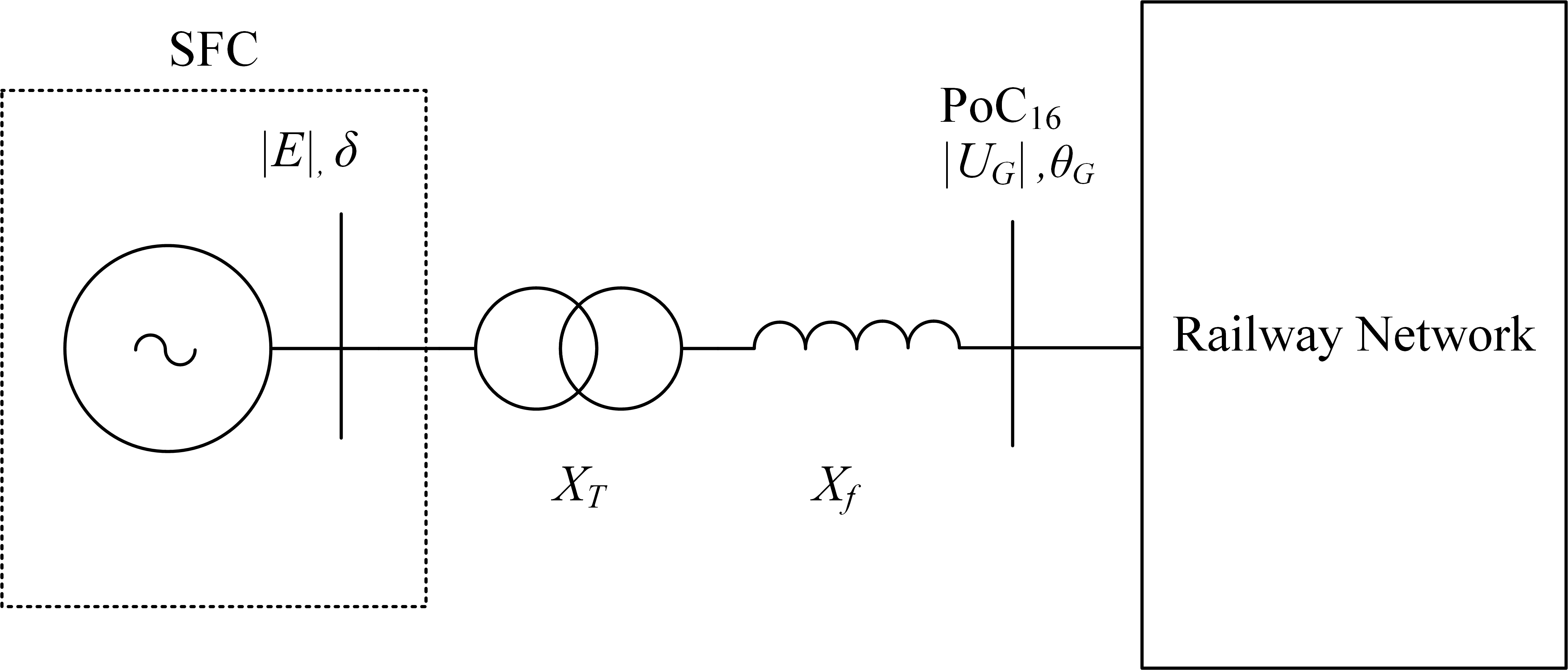} 
\caption{Electromechanical SFC inverter model.}\label{Fig:SFC_circuit_simple}
\end{figure}

The band-stop filter is neglected in the SFC inverter model for the illustrative studies shown later, resulting in a simplified connection to the railway grid as seen in \Cref{Fig:SFC_circuit_simple}. 

Note that all equations in this paper are expressed in the p.u. system.

\subsection{Steady-state model}
In synchronous railway grids, the SFC inverter is controlled to mimic the steady-state phase angle and voltage behaviour of an RFC at the $\text{PoC}_{16}$. The result is that it allows parallel operation in steady-state between RFC and SFC in a converter station \cite{Abrahamsson117,olofsson1996}.

\subsubsection{Voltage phase angle}
A voltage phase shift is introduced when active power flows through an RFC. The voltage phase shift for an RFC is the sum of load angle from the three-phase synchronous motor and single-phase synchronous generator \cite{Abrahamsson117,olofsson1996}, and is
\begin{align}
\psi = \frac{1}{3} \arctan\left(\frac{X_{q}^{\text{m}}\cdot P_D^{\text{m}}}{{|U^{\text{m}}|}^2 + X_{q}^{\text{m}} Q_D^{\text{m}}}\right)
+ \arctan \left(\frac{X_{q}^{\text{g}}P_G^{\text{g}}}{{|U^{\text{g}}|}^2 + X_{q}^{\text{g}} Q_G^{\text{g}}}\right)  \label{Eq 1: Phaseshift},
\end{align}
where $X_q$ is the quadrature reactance, $|U|$ is the voltage magnitude, $P$ and $Q$ are the active power and reactive power , respectively. The superscripts "m" and "g" stand for motor and generator respectively and the and the subscripts "G" and "D" stand for generation and demand (i.e.load/consumption), respectively. 

For simplicity the step-down transformer is included in the quadrature reactance of the motor, whereas the step-up transformer is included in the quadrature reactance of the generator. 

It is assumed that the RFC is lossless, which results in the steady-state demanded power of the motor, $P_D^{\text{m}}$, to be equal to the generated power, $P_G^{\text{g}}$, of the generator. When the RFC motor consumes active power, then the RFC generator produces active power. The resulting terminal phase angle $\theta^{\text{g}}$ of an RFC generator at the railway side relative to the voltage phase angle $\theta_{50}$ of the public grid where the RFC motor is connected to is
\begin{equation}
\theta^{\text{g}} = \frac{\theta_{50}}{3} - \psi. \label{Eq 2: Termimal angle}
\end{equation} 
Note that the terminal phase angle of an RFC generator is expressed in $16\frac{2}{3}$~Hz angles, therefore the division with three of $\theta_{50}$.

\subsubsection{Voltage magnitude}
The voltage magnitude after the step-up transformer of an RFC generator is controlled such way that with increased reactive power injection, the voltage magnitude will fall after the step-up transformer.
To allow for parallel operation, the same requirement is set on the steady-state voltage control for the SFC. The voltage after the step-up transformer, for both the RFC and the SFC, is given by the following expression:
\begin{equation}
|U^{\text{g}}| = |U_0| - K_UQ_G^{\text{g}} \label{Eq: Droop control}
\end{equation}
where $|U_0|$ is the no-load voltage and $K_U$ is the droop coefficient scaled according to the rating of the SFC or RFC.

%
%


\subsection{SFC dynamic model}
As the model is for electromechanical stability studies, a first-order transfer function can be used to model the internal circuit dynamic of a converter \cite{HVDCbook2016,Machowski2008}. The SFC inverter internal dynamics are therefore approximated with a first-order transfer function with the time constant $T_c$. The focus is then on the control of the SFC inverter voltage magnitude and phase angle.

\subsubsection{Phase angle control}
To control the voltage phase angle $\theta^{\text{g}}$ so that it follows \Cref{Eq 2: Termimal angle} at the $\text{PoC}_{16}$, the SFC inverter will regulate its voltage phase angle $\delta$.

An SFC inverter allows generating any required voltage from a DC voltage independent of the strength of the supplying three-phase grid. As it assumed that the SFC rectifier is connected to an infinite bus, the denominator in \Cref{Eq 1: Phaseshift} for the motor is set to unity.


\begin{figure}
\centering
\includegraphics[scale=0.8]{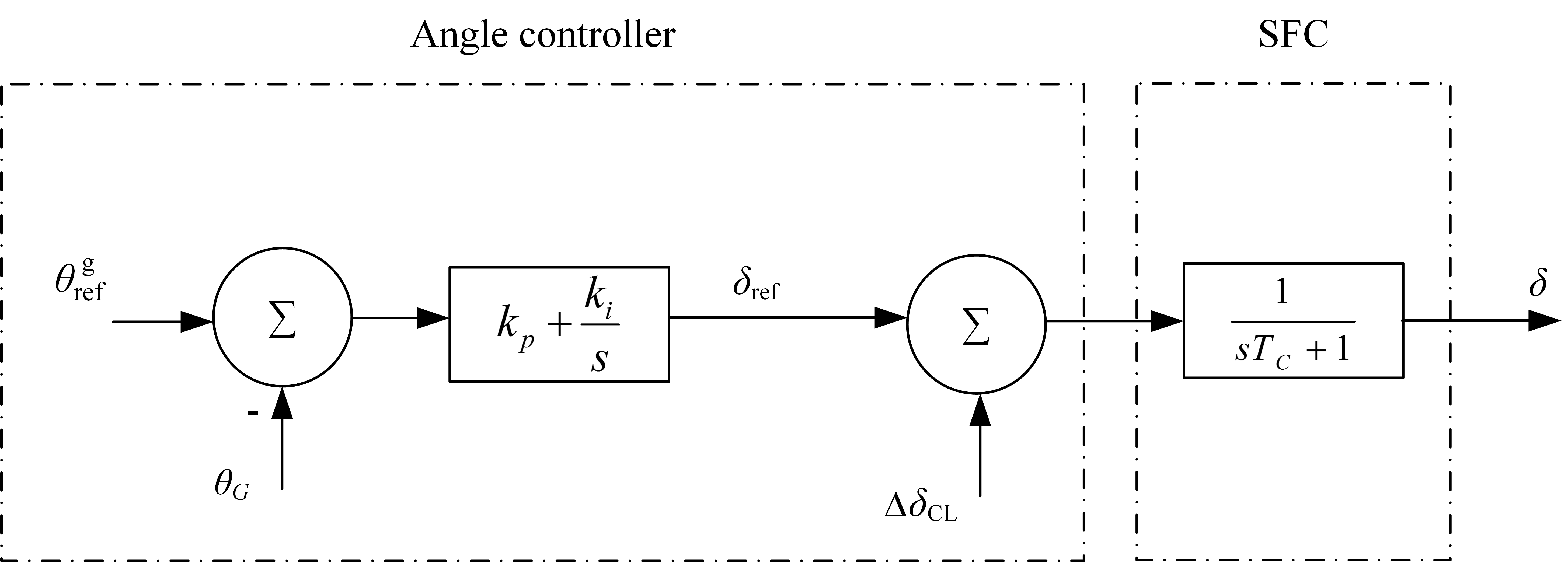} 
\caption{Angle controller.}\label{Fig:SFC_Angle}
\end{figure}

Measuring voltage $|U_G|$, active power $P_G$ and reactive power $Q_G$ at the $\text{PoC}_{16}$, the voltage phase angle reference $\theta_{\text{ref}}^{\text{g}}$ is calculated as 
\begin{equation}
\theta_{\text{ref}}^{\text{g}} =\frac{\theta_{50}}{3} - \frac{1}{3} \arctan(X_{q}^{\text{m}}\cdot P_G) - \arctan\left(\frac{X_{q}^{\text{g}}\cdot P_G}{{|U_G|}^2 + X_{q}^{\text{g}} Q_G}\right) \label{Eq : angle reference}
\end{equation}  
It should be observed that in \Cref{Eq : angle reference} the RFC transformer reactances are included in the quadrature reactances of the motor and generator, respectively. 

The phase angle control is implemented with a PI regulator, its inputs are the measured phase angle $\theta_G$ at the $\text{PoC}_{16}$ and the calculated $\theta_{\text{ref}}^{\text{g}}$ from \Cref{Eq : angle reference}. The output of the PI regulator is the SFC voltage phase angle reference $\delta_{\text{ref}}$ and is
\begin{equation}
\delta_{\text{ref}} = K_p(\theta_{\text{ref}}^{\text{g}}  - \theta_G) + K_i \int_{t_0}^{t} (\theta_{\text{ref}}^{\text{g}} - \theta_G) dt.
\end{equation}
When current limitation is active, see \Cref{subCL}, the signal $\Delta\delta_{\text{CL}}$ from the current limitation control is added to $\delta_{\text{ref}}$. The resulting signal $\delta_{\text{ref}}+\Delta\delta_{\text{CL}}$ is sent to the SFC inverter, where $\delta$ is the voltage phase angle output of the inverter, see \Cref{Fig:SFC_Angle}.

%

\subsubsection{Voltage control}
\begin{figure}
\centering
\includegraphics[scale=0.8]{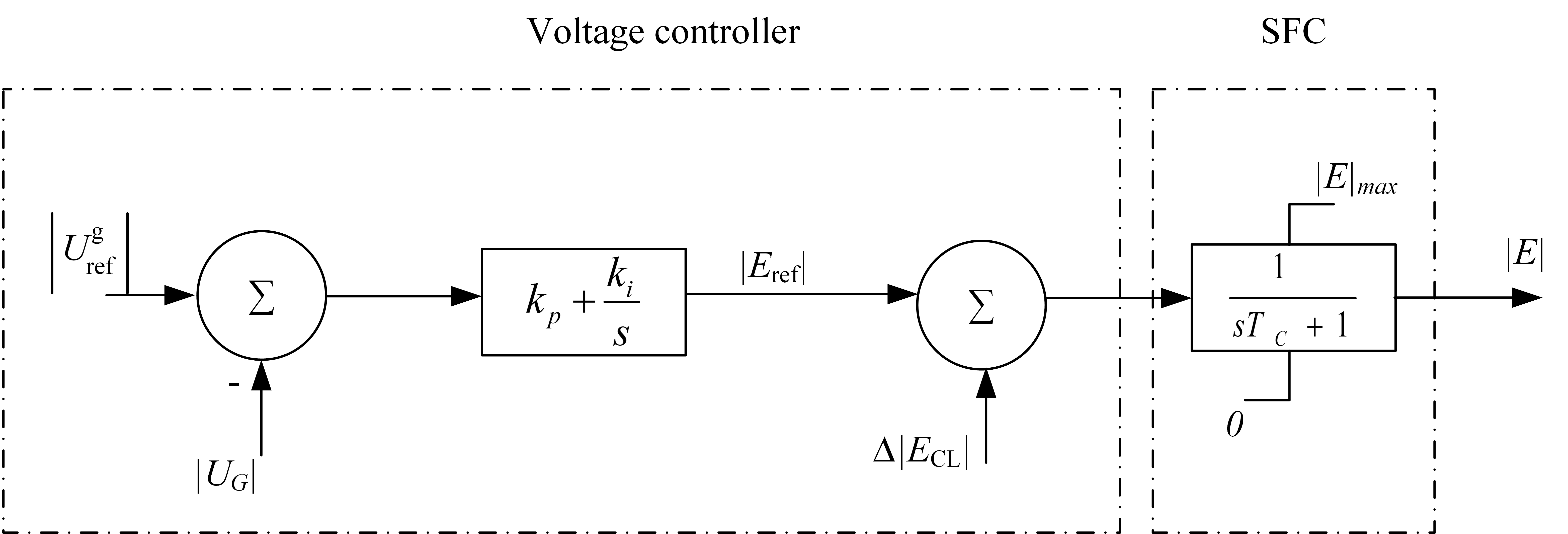} 
\caption{Voltage controller.}\label{Fig:SFC_volt}
\end{figure}

Measuring the reactive power at the $\text{PoC}_{16}$, the voltage magnitude reference $|U^{\text{g}}_\text{ref}|$ is calculated by using \Cref{Eq: Droop control}. The inputs to the voltage controller is the calculated $|U^{\text{g}}_\text{ref}|$ and the measured voltage $|U_G|$ at the $\text{PoC}_{16}$.
The voltage controller is implemented with a PI regulator. The output of the PI regulator is the SFC voltage magnitude reference $|E_\text{ref}|$ and is
\begin{equation}
|E_\text{ref}| = K_p(|U^{\text{g}}_\text{ref}| - |U_G|) + K_i \int_{t_0}^{t} (|U^{\text{g}}_\text{ref}| - |U_G|) dt.
\end{equation}
If current limitation is active, see \Cref{subCL}, an additional signal $\Delta |E_\text{CL}|$ from the current limitation control is added to $|E_\text{ref}|$. The resulting signal $|E_\text{ref}| + \Delta |E_\text{CL}|$ is transferred to the SFC inverter that gives the final inverter voltage output $|E|$, see \Cref{Fig:SFC_volt}. The inverter output voltage is limited to $|E|_{max}$: the maximum voltage the SFC inverter can generate when its modulation index is equal to unity. In this model the maximum inverter voltage $|E|_{max}$ is set to 1.15 p.u.


\subsubsection{Current limitation control} \label{subCL}
As power electronics cannot handle large currents above their rated current, converters are equipped with control systems that limits the current to a given value that will not damage the converter. The modelling of current limitation control for the SFC inverter has been inspired by the current limitation used in VSC-HVDC control in \cite{Simpow2010}.

The SFC inverter current through the transformer and filter is
\begin{equation}
\abs{I}\text{e}^{j\gamma} = \frac{\abs{E}\textit{e}^{j\delta}-\abs{U_G}\textit{e}^{j\theta_G}}{j(X_T + X_f)}, \label{eq1cl}
\end{equation}
and the magnitude of the current is
\begin{equation}
\abs{I} = \frac{\abs{E}\textit{e}^{j(\delta-\gamma)}-\abs{U_G}\textit{e}^{j(\theta_G -\gamma)}}{j(X_T + X_f)}. \label{eq2cl}
\end{equation}
The left hand side of \Cref{eq2cl} is a real value, and therefore the imaginary part of the right hand side of \Cref{eq2cl}  has to be zero. Thus, the real part of \Cref{eq2cl} is
\begin{equation}
|I| = \frac{|E|\sin(\delta-\gamma) - |U_G|\sin(\theta_G-\gamma)}{{(X_T + X_f)} }. \label{eq3cl}
\end{equation}
Assuming small changes in the voltage magnitude and phase angle at the $\text{PoC}_{16}$, the railway grid can be seen as a constant voltage \cite{Simpow2010}. 
Differentiation with respect to $\abs{E}$ and $\delta$ results in:
\begin{align}
\frac{d\abs{I}}{d\abs{E}} &= \frac{\sin(\delta - \gamma)}{(X_T + X_f)}  \label{lim1}\\
\frac{d\abs{I}}{d\delta} &= \frac{\abs{E}\cos(\delta - \gamma)}{(X_T + X_f)}  \label{lim2}.
\end{align}
The change of current magnitude as a function of changes in voltage magnitude and phase of the SFC can be expressed as
\begin{equation}
\Delta \abs{I} = \frac{\sin(\delta - \gamma)}{(X_T + X_f)} \Delta \abs{E} + \frac{\abs{E} \cos(\delta - \gamma)}{(X_T + X_f)} \Delta \delta \label{Eq: CLimit}.
\end{equation}

A PI regulator is used to reduce the measured inverter current magnitude $|I|$ when it exceeds $|I|_{max}$. The current limitation control is only active when $|I| > |I|_{max}$. Note that the values of the reactances are included in proportionality constant of the PI regulator. 

\begin{figure}
\centering
\includegraphics[scale=0.7]{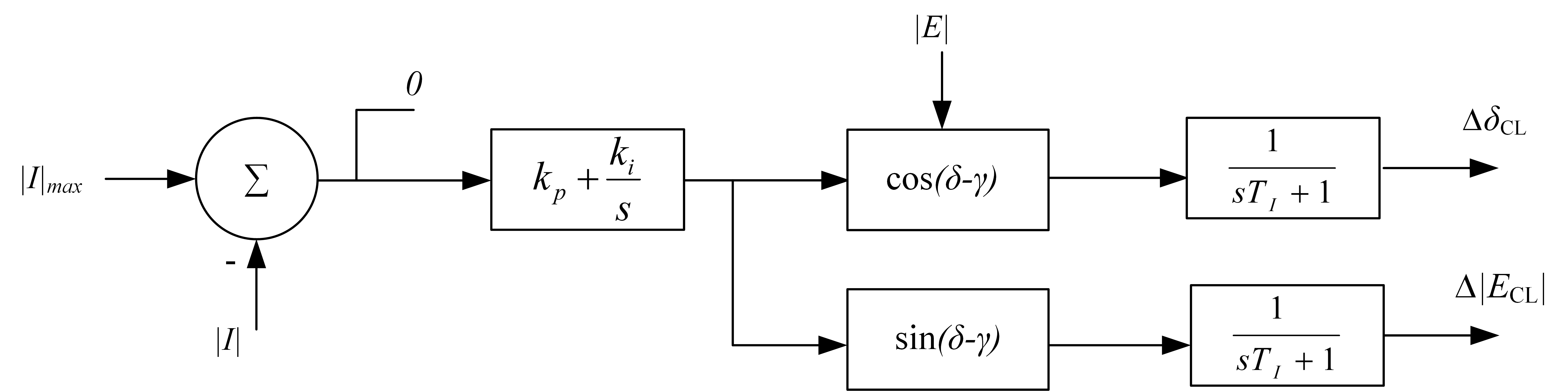} 
\caption{Current limitation controller.}\label{Fig:Curren_limit}
\end{figure}

The signals from the PI regulator are weighted with $\sin(\delta_{SFC} - \gamma)$ and $\abs{E} \cos(\delta - \gamma)$, respectively. The signals are the filtered through a first-order filter with the time constant $T_I$ as seen in \Cref{Fig:Curren_limit}. The outputs of the current limitation controller are $\Delta |E_\text{CL}|$ and $\Delta\delta_{\text{CL}}$. The changes in voltage magnitude according to \Cref{Eq: CLimit} are added to the ouputs of the value of voltage magnitude and phase angle controller, respectively, see \Cref{Fig:SFC_Angle,Fig:SFC_volt}.
The integrator of the PI regulator of the current limitation is reset to zero when $|I|<|I|_{max}$.

As shown by \Cref{Eq: CLimit}, the angle between voltage and current will decide the amount of active and reactive power supplied by the SFC inverter during current limitation.  


\section{Implementation} \label{Sec: Implemenation}
Dynamics of the railway grid and the public grid is not considered as the SFC model presented is intended for electromechanical stability studies, and a quasi-steady-state description of the system is used.


The model has been implemented in MatLab Simulink where the variable step solver \textit{ode45} has been used. The steady-state values are found through a load-flow study using a part of the software TPSS developed in \cite{Abrahamsson117}.

For the simulations presented in \Cref{Sec: Validation}, the SFC inverter model is connected to the grid as an equivalent Norton current source. The loads are modelled as constant shunt admittances for the dynamic simulation. The value of the load admittance is based on results obtained from load-flow solution.

The PI regulators of the SFC inverter have been manually tuned to fit the RMS phasor characteristics of voltage and current from the measured data. The data of the system is given in \Cref{Tab: System data,Tab: SFC transformer,Tab: SFC control,Tab: SFC current controll}.

\begin{table}
\centering
\begin{tabular}{ l | c }
	\hline			
	Base Power & 10 [MVA] \\
	Base Voltage & 16.5 [kV] \\
	\hline  
\end{tabular}
\caption{System data.} \label{Tab: System data}
\end{table}

\begin{table}
\centering
\begin{tabular}{ l | c }
	\hline			
	Transformer Rating  & 17.4 [MVA] \\
	Leakage Reactance & 16.65 \%  \\
	Secondary Voltage & 16.5 [kV]\\
	Inductive filter, $X_f$ & 32 [mH]\\
	\hline  
\end{tabular}
\caption{SFC transformer parameters and inductive filter parameter} \label{Tab: SFC transformer}
\end{table}

\begin{table}
\centering
\begin{tabular}{ l | c }
	\hline			
	Quadrature Reactance Motor$X^m_{q}$ & 0.49 [p.u.] \\
	Motor Transformer Leakage Reactance & 7.9 \% \\
	Rated Power Motor Transformer & 10.7 [MVA]\\
	Nominal Voltage Motor & 6.3 [kV] \\
	Rated Power Motor & 10.7 [MVA]\\
	Quadrature Reactance Generator $X^g_{q}$ & 0.53 [p.u.] \\
	Generator Transformer Leakage Reactance & 4.2 \% \\
	Nominal Voltage Generator & 5.2 [kV] \\
	Transformer Primary Voltage & 5.2 [kV] \\
	Transformer Secondary Voltage & 17 [kV] \\
	Rated Power Generator Transformer & 10 [MVA]\\
	Rated Power Generator  & 10 [MVA]\\
	\hline  
\end{tabular}
\caption{RFC parameter for SFC control.} \label{Tab: SFC control}
\end{table}

\begin{table}
\centering
\begin{tabular}{ l | c }
	\hline			
	Voltage/Angle Regulator & $k_p=0.02$, $k_i=2$  \\
	Voltage droop $K_U$ & 3\% \\
	Converter dynamics $T_C$ & 50 [ms]\\
	Current limitation regulator & $k_p=12$, $k_i=75$\\
	Current limitation time constant $T_I$ & 0.02 [ms]\\
	\hline  
\end{tabular}
\caption{SFC control paramteres.} \label{Tab: SFC current controll}
\end{table}
\section{Validation} \label{Sec: Validation}
\subsection{Measuring system}
The Swedish Transport Administration (TrV) have installed measuring instruments at the railway-side of several frequency converter units. The measurements units are power-quality monitors adapted to 16$\frac{2}{3}$~Hz and compliance with Class A under IEC 61000-4-30 \cite{IEC61000430}. Both monitors and the analysis software are delivered by Metrum Sweden AB. Voltage and current at the $\text{PoC}_{16}$ from the SFC inverter are measured in time-domain. The measured data is transferred to a database, from which the user via dedicated software can, for example, select measured voltages and current waveforms for events during which certain pre-set thresholds are exceeded.

The measured data for the selected event is from a single SFC inverter operating in a converter station; the other converters in that station were disconnected from the railway grid during the measured disturbances. The converter station is feeding an single catenary section. The measured data is transformed to RMS form by the analysis software provided Metrum Sweden AB.

%

\subsection{Cases}
\begin{figure}
\centering
\includegraphics[scale=0.8]{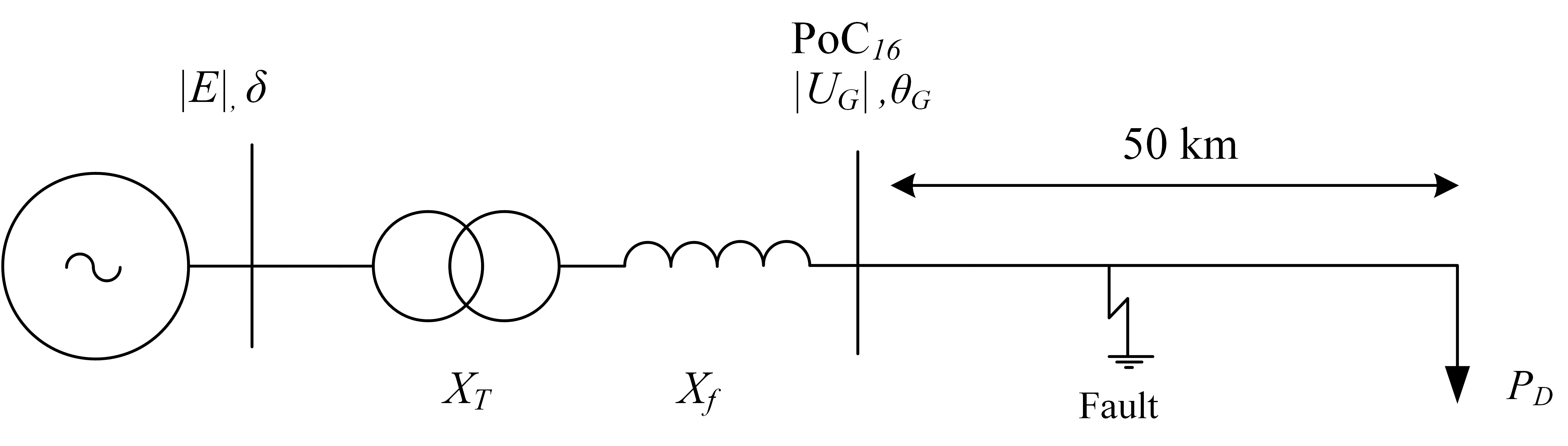} 
\caption{Simulated system.}\label{Fig:simulation}
\end{figure}

The SFC inverter is simulated in single-feeding mode, see \Cref{Fig:simulation}, as the measured SFC is feeding one catenary section. The catenary system used in the simulations is an Auto Transformer (AT) catenary system, as the measured SFC is connected to an AT system. The equivalent impedance of an AT catenary system is $Z_{\text{AT}} = 0.0335+j0.031 \frac{\Omega}{\text{km}}$, with an initial impedance of $Z_\text{init} = 0.189+j0.293$ at the $\text{PoC}_{16}$ \cite{Abrahamsson117,Fridman2006}.   

The SFC inverter is simulated for four cases where a fault, $Z_{fault}$, from the overhead contact line to ground is applied. The values of the pre-fault load and $Z_{fault}$  have been selected for the different cases to match the measured pre-fault and during-fault RMS voltages and currents. The faults of the four cases are cleared at different time instance, based on the measured data. The threshold used to determine when current limitation is active is when the measured current is above $|I|_{max}=2.0$ p.u.


The cases that are used for validation for reproducing the measured data are:

\begin{itemize}	
\item Case 1: 60 ms, current limitation not active,  $Z_{fault} = 0.47 +j0.15$ p.u. 25 km from $\text{PoC}_{16}$ of the SFC inverter. Train Load $P_{D} = 2.4$ MW.
\item Case 2: 80 ms, current limitation not active, $Z_{fault} = 0.5 + j0.2$ p.u. 25 km from $\text{PoC}_{16}$ of SFC inverter.
Train Load $P_{D} = 2.75$ MW.
\item Case 3: 120 ms, current limitation active, $Z_{fault} = 0 + j0.13$ p.u. 15 km from $\text{PoC}_{16}$ of SFC inverter.
Train Load $P_{D} = 4.25$ MW.
\item Case 4: 270 ms, current limitation active, $Z_{fault} = 0 + j0.15$ p.u. 20 km from $\text{PoC}_{16}$ of SFC inverter.
Train Load $P_{D} = 1.5$ MW.
\end{itemize} 

\subsection{Current limitation not active}
\begin{figure}[h!]
\centering
\setlength\fheight{6cm}
\setlength\fwidth{\textwidth}
\input{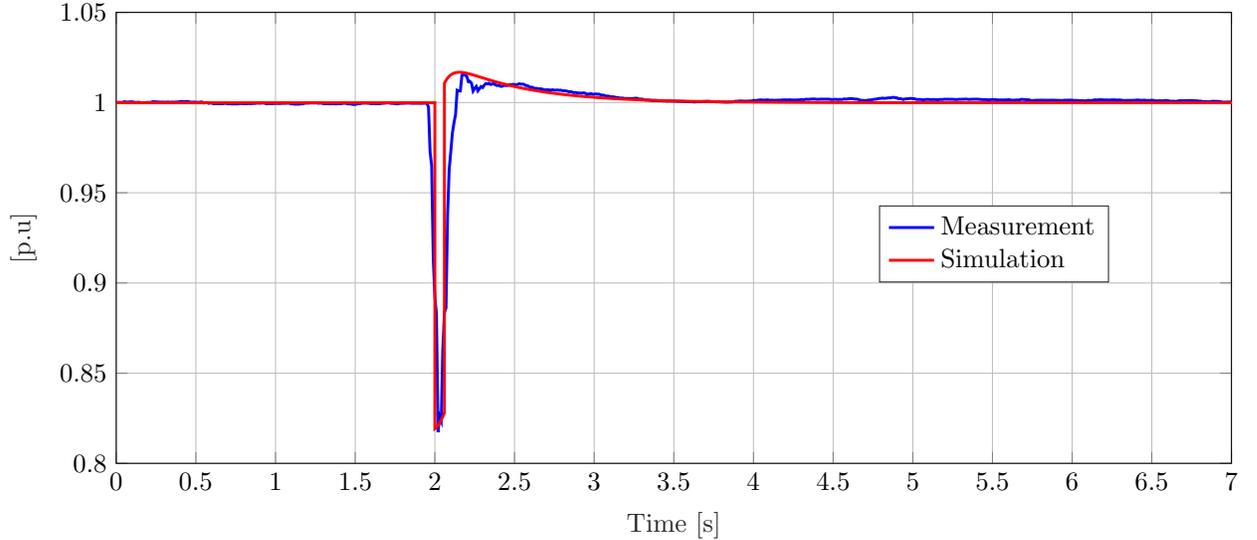} 
\caption{Case 1: Voltage [p.u.]} \label{Fig: Case1 Voltage}
\end{figure}

\begin{figure}[h!]
\centering
\setlength\fheight{6cm}
\setlength\fwidth{\textwidth}
\input{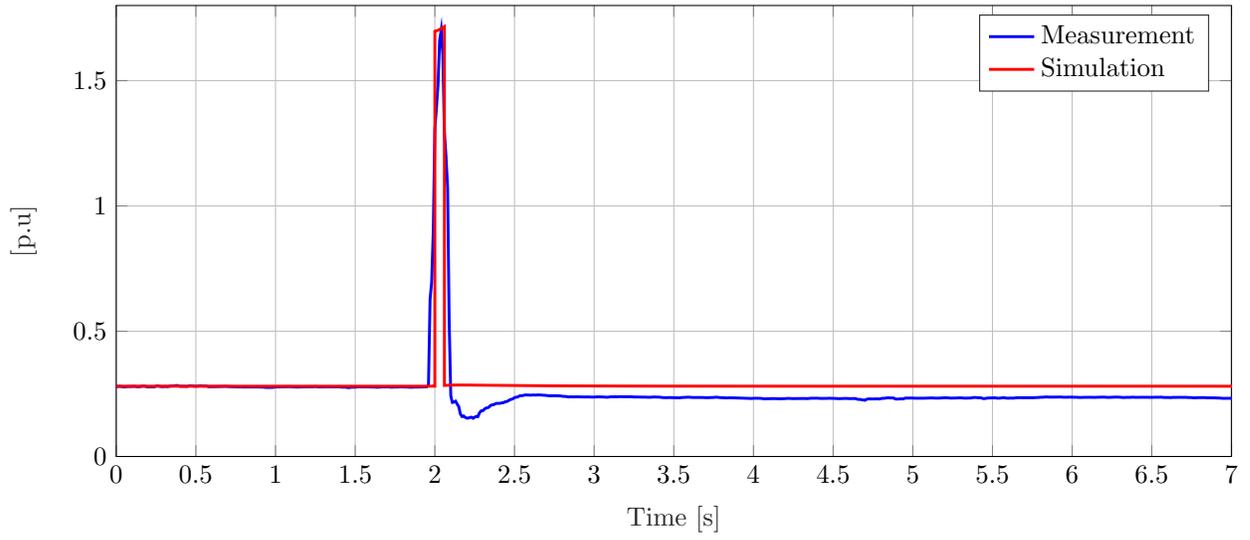} 
\caption{Case 1: Current [p.u.]} \label{Fig: Case1 Current}
\end{figure}

\begin{figure}[h!]
\centering
\setlength\fheight{6cm}
\setlength\fwidth{\textwidth}
\input{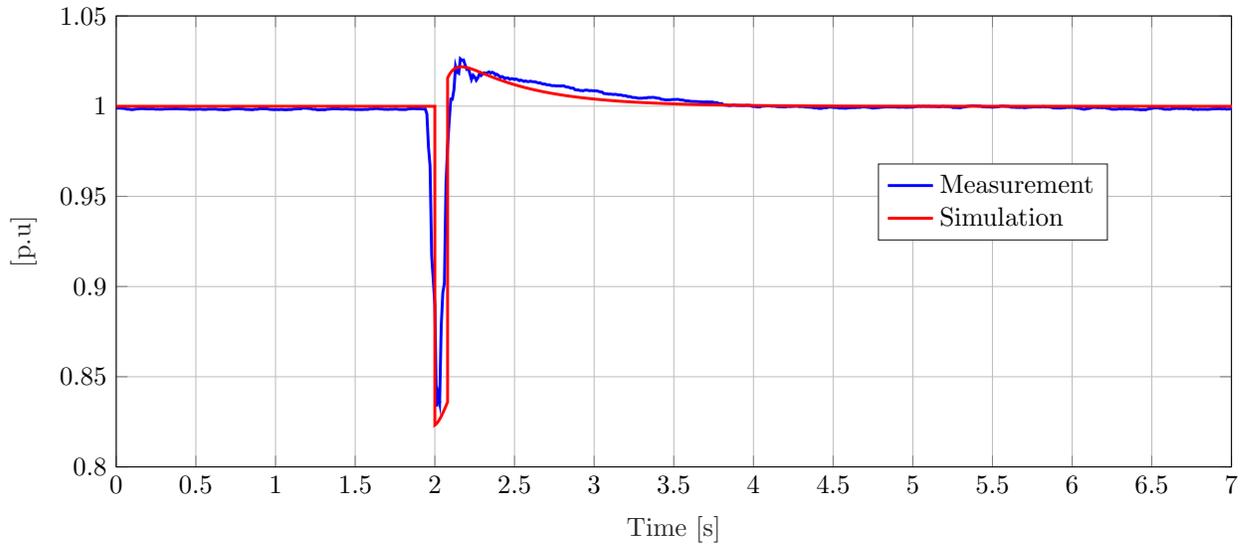} 
\caption{Case 2: Voltage [p.u.]} \label{Fig: Case2 Voltage}
\end{figure}

\begin{figure}[h!]
\centering
\setlength\fheight{6cm}
\setlength\fwidth{\textwidth}
\input{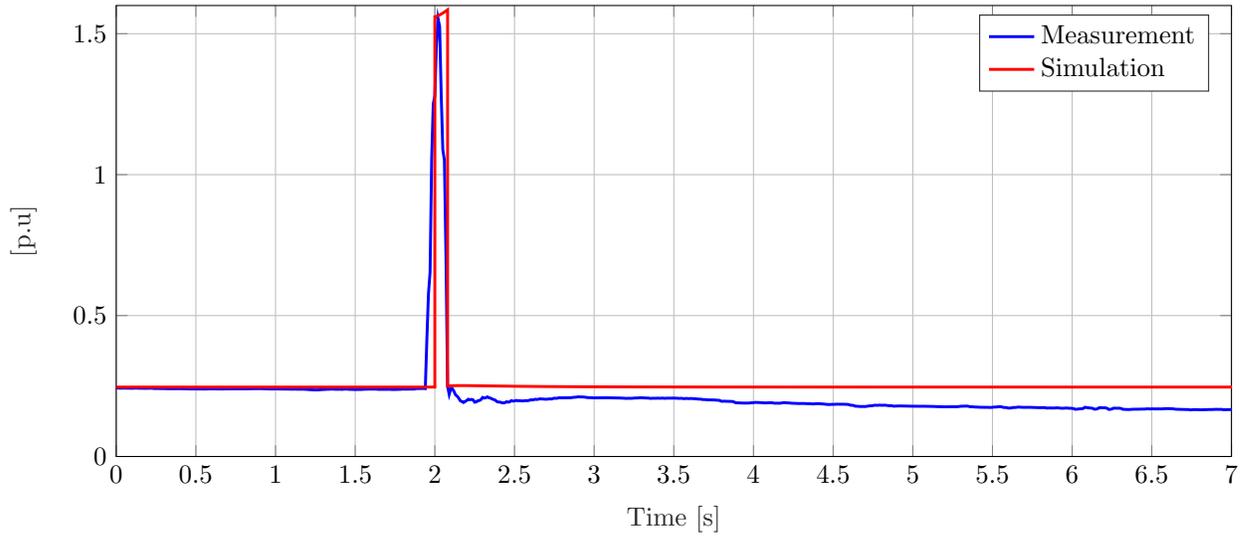} 
\caption{Case 2: Current [p.u.]} \label{Fig: Case2 Current}
\end{figure}

When the system is subjected to a fault, the proposed model reproduces the measured data adequately both in voltage and current as seen for Case 1 (\Cref{Fig: Case1 Voltage,Fig: Case1 Current}) and Case 2 (\Cref{Fig: Case2 Voltage,Fig: Case2 Current}). The undershoot in currents in post-fault from the measured data may origin from the train, which is not captured by the load model used. 

\subsection{Current limitation active}

\begin{figure}[h!]
\centering
\setlength\fheight{6cm}
\setlength\fwidth{\textwidth}
\input{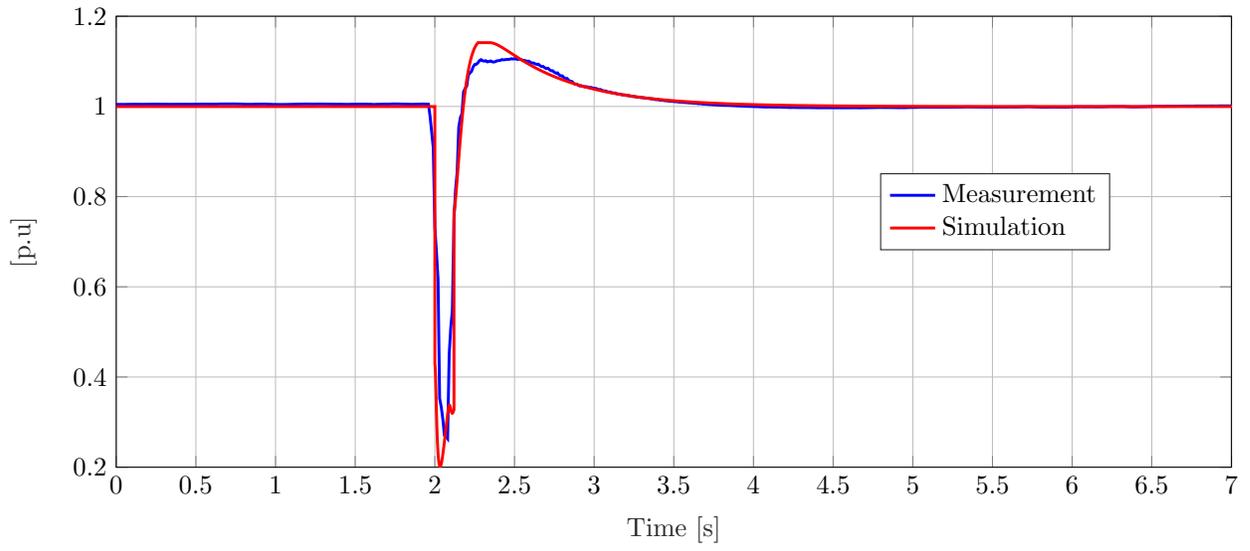} 
\caption{Case 3: Voltage [p.u.]} \label{Fig: Case3 Voltage}
\end{figure}

\begin{figure}[h!]
\centering
\setlength\fheight{6cm}
\setlength\fwidth{\textwidth}
\input{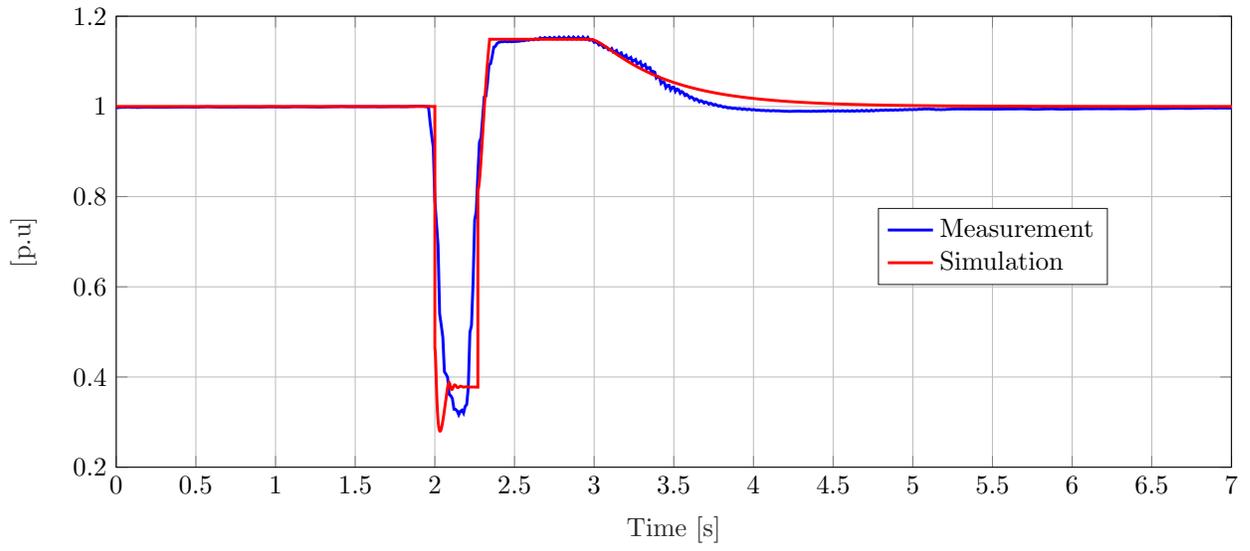} 
\caption{Case 4: Voltage [p.u.]} \label{Fig: Case4 Voltage}
\end{figure}

The measured voltages in \Cref{Fig: Case3 Voltage,Fig: Case4 Voltage} (Case 3 and Case 4, respectively) has large overshoots. It may indicate that the voltage control of the measured SFC inverter has no anti-windup scheme applied, and therefore the SFC model has been implemented with no anti-windup on the voltage control. 

The SFC model simulated reproduces the measured RMS voltages with an acceptable accuracy in Case 3 and Case 4. There are difference in voltage levels in Case 4 between measured and simulated voltage during current limitation (2.10 seconds to 2.27 seconds). The simulated voltage is higher than the measurement which results that the windup in the control of the voltage is less in the simulated case compared to the measured voltage after fault clearance. The behaviour of the voltage during fault in the simulated case is reasonable given that the fault impedance and the load impedance is constant. A plausible explanation why the voltage continues to drop in the measurement is that the fault has a variable impedance. 

\begin{figure}[h!]
\centering
\setlength\fheight{6cm}
\setlength\fwidth{\textwidth}
\input{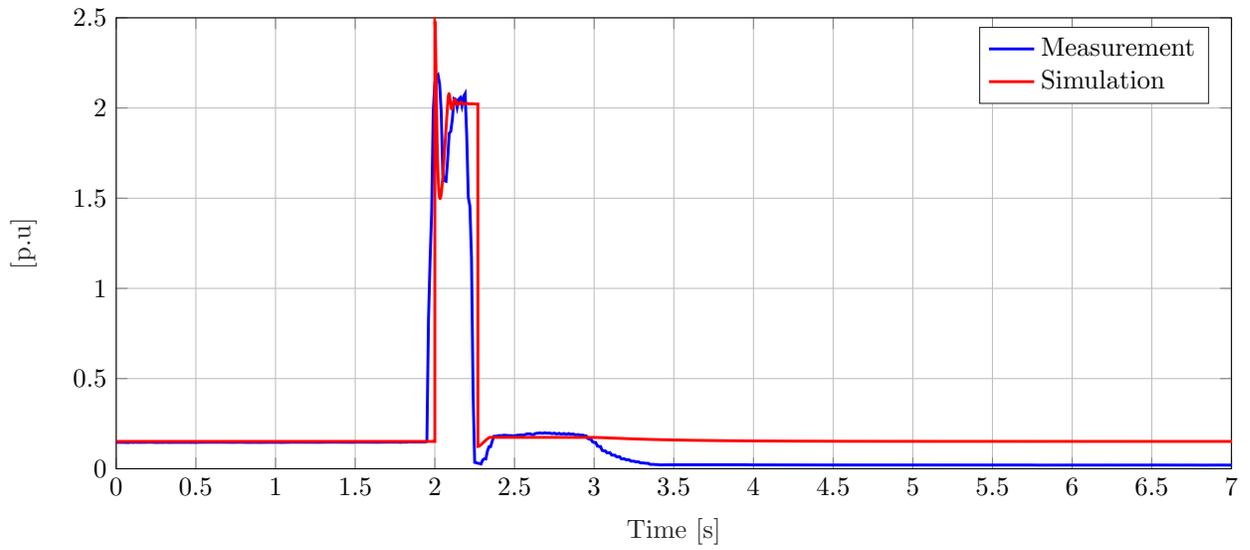} 
\caption{Case 4: Current [p.u.]} \label{Fig: Case4 Current}
\end{figure}

The SFC inverter model reproduces with an acceptable accuracy the measured current in Case 4 during the fault as seen in \Cref{Fig: Case4 Current}. The measured current in Case 3 seems to have sub-cycles changes which cannot be reproduced with the proposed SFC model as the simulation is done in RMS phasor, where only the fundamental frequency is considered. Despite the aforementioned discrepancies, the proposed SFC model reproduces some of the most important characteristics of the measured current during fault.

Note that in Case 4 after 3 seconds in \Cref{Fig: Case4 Current} the measured current of the SFC inverter is reduced to zero. A plausible explanation is that an SFC inverter protection system or another railway protection system is activated which disconnects the SFC inverter.

%

%

\begin{figure}[h!]
\centering
\setlength\fheight{6cm}
\setlength\fwidth{\textwidth}
\input{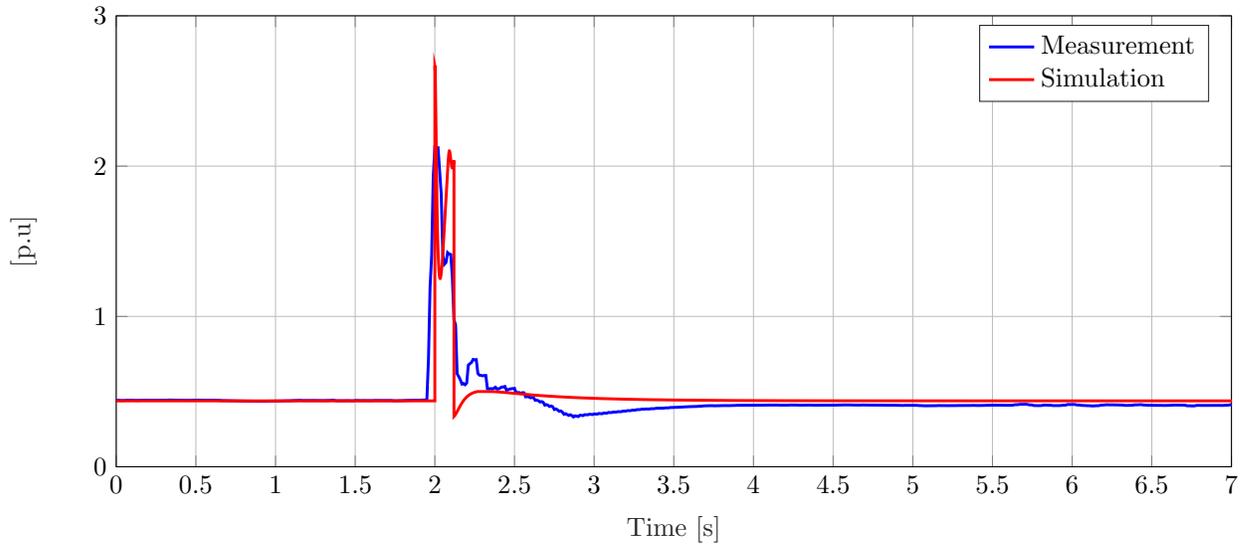} 
\caption{Case 3: Current [p.u.]} \label{Fig: Case3 Current}
\end{figure}

\clearpage
\section{Conclusion} \label{Sec: Conclusions}
This paper has presented an open SFC model that is adequate for electromechanical simulations and stability studies for synchronous low-frequency railway grids of 16$\frac{2}{3}$~Hz.

Simulations in the phasor domain have validated the model against measured RMS magnitude of voltage and current of an SFC inverter operating in the Swedish railway grid. Comparing the measured data with the simulated results, it can be concluded that the model adequately reproduces the measured RMS voltages and currents. Windup in the voltage of the measured SFC inverter are properly reproduced with the proposed SFC model.  Thus, it can be concluded that the  SFC model proposed provides an adequate description of the main characteristics of an SFC inverter behaviour, both when current limitation is inactive and when it is active.


Several simplifications were made in the SFC model and its implementation. For most of these simplifications, the SFC model can easily be extended with more functions and details for improved accuracy.


The model itself is not limited to synchronous low-frequency railways grid. With changes in control objective of the SFC, it can be used for electromechanical stability studies in the phasor domain for three phase systems.

The simplicity of the SFC model presented allows for easy implementation in power system softwares or numerical softwares such as MatLab, for electromechanical stability studies of synchronous railway grids or three-phase grids. 

\section*{Acknowledgements}
The financial support for this project from the Swedish Transport Administration is greatly acknowledged.


\small{
\bibliographystyle{IEEEtran}
\bibliography{library}

\begin{thebibliography}{10}
\providecommand{\url}[1]{#1}
\csname url@samestyle\endcsname
\providecommand{\newblock}{\relax}
\providecommand{\bibinfo}[2]{#2}
\providecommand{\BIBentrySTDinterwordspacing}{\spaceskip=0pt\relax}
\providecommand{\BIBentryALTinterwordstretchfactor}{4}
\providecommand{\BIBentryALTinterwordspacing}{\spaceskip=\fontdimen2\font plus
\BIBentryALTinterwordstretchfactor\fontdimen3\font minus
  \fontdimen4\font\relax}
\providecommand{\BIBforeignlanguage}[2]{{%
\expandafter\ifx\csname l@#1\endcsname\relax
\typeout{** WARNING: IEEEtran.bst: No hyphenation pattern has been}%
\typeout{** loaded for the language `#1'. Using the pattern for}%
\typeout{** the default language instead.}%
\else
\language=\csname l@#1\endcsname
\fi
#2}}
\providecommand{\BIBdecl}{\relax}
\BIBdecl

\bibitem{Steimel2012}
\BIBentryALTinterwordspacing
A.~Steimel, ``{Power-electronic grid supply of AC railway systems},'' in
  \emph{2012 13th International Conference on Optimization of Electrical and
  Electronic Equipment (OPTIM)}.\hskip 1em plus 0.5em minus 0.4em\relax IEEE,
  may 2012, pp. 16--25. [Online]. Available:
  \url{http://ieeexplore.ieee.org/lpdocs/epic03/wrapper.htm?arnumber=6231844}
\BIBentrySTDinterwordspacing

\bibitem{Lars2012}
L.~Abrahamsson, S.~{\"{O}}stlund, and T.~Sch{\"{u}}tte, ``{Use of converters
  for feeding of AC railways for all frequencies},'' \emph{Energy for
  Sustainable Development,}, vol.~16, no. nr 3, pp. 368--378, 2012.

\bibitem{steimel2008}
A.~Steimel, \emph{{Electric Traction - Motive Power and Energy Supply}},
  E.~Krammer, Ed.\hskip 1em plus 0.5em minus 0.4em\relax Munich: Oldenbourg
  Industrieverlag GmbH, 2008.

\bibitem{tomp1898}
\BIBentryALTinterwordspacing
S.~P. Thompson and M.~Field, ``{Rotatory converters},'' \emph{Journal of the
  Institution of Electrical Engineers}, vol.~27, no. 137, pp. 651--689, nov
  1898. [Online]. Available:
  \url{http://digital-library.theiet.org/content/journals/10.1049/jiee-1.1898.0028}
\BIBentrySTDinterwordspacing

\bibitem{Pfeiffer1997}
\BIBentryALTinterwordspacing
A.~Pfeiffer, W.~Scheidl, M.~Eitzmann, and E.~Larsen, ``{Modern rotary
  converters for railway applications},'' in \emph{Proceedings of the 1997
  IEEE/ASME Joint Railroad Conference}.\hskip 1em plus 0.5em minus 0.4em\relax
  IEEE, 1997, pp. 29--33. [Online]. Available:
  \url{http://ieeexplore.ieee.org/lpdocs/epic03/wrapper.htm?arnumber=581349}
\BIBentrySTDinterwordspacing

\bibitem{Abrahamsson117}
L.~Abrahamsson, ``{Railway Power Supply Models and Methods for Long-term
  Investment Analysis},'' Licentiate Thesis, KTH, Electric Power Systems, 2008.

\bibitem{Olofsson1995}
\BIBentryALTinterwordspacing
M.~Olofsson, ``{Optimal operation of the Swedish railway electrical system},''
  in \emph{International Conference on Electric Railways in a United Europe},
  vol. 1995.\hskip 1em plus 0.5em minus 0.4em\relax IEE, 1995, pp. 64--68.
  [Online]. Available:
  \url{http://digital-library.theiet.org/content/conferences/10.1049/cp{\_}19950179}
\BIBentrySTDinterwordspacing

\bibitem{Ostlund2012}
S.~{\"{O}}stlund, \emph{{Electric Railway Traction}}.\hskip 1em plus 0.5em
  minus 0.4em\relax Stockholm: School of Electrical Engineering, Royal Insitute
  of Technology, 2012.

\bibitem{Eitzmann1997}
\BIBentryALTinterwordspacing
M.~Eitzmann, J.~Paserba, J.~Undrill, C.~Amicarella, A.~Jones, E.~Khalafalla,
  and W.~Liverant, ``{Model development and stability assessment of the Amtrak
  25 Hz traction system from New York to Washington DC},'' in \emph{Proceedings
  of the 1997 IEEE/ASME Joint Railroad Conference}.\hskip 1em plus 0.5em minus
  0.4em\relax IEEE, 1997, pp. 21--28. [Online]. Available:
  \url{http://ieeexplore.ieee.org/lpdocs/epic03/wrapper.htm?arnumber=581348}
\BIBentrySTDinterwordspacing

\bibitem{HVDCbook2016}
E.~Mircea, L.~Chen‐Ching, and A.~Edris, Eds., \emph{{Advanced Solutions in
  Power Systems: HVDC, FACTS, and Artificial Intelligence}}.\hskip 1em plus
  0.5em minus 0.4em\relax New Jersey: John Wiley {\&} Sons, Inc, 2016.

\bibitem{JingLiu2009}
\BIBentryALTinterwordspacing
J.~{Jing Liu}, M.~{Minxiao Han}, and X.~{Xiuyu Chen}, ``{Development of VSC
  HVDC quasi-steady-state model and its application},'' in \emph{2009
  International Conference on Sustainable Power Generation and Supply}.\hskip
  1em plus 0.5em minus 0.4em\relax IEEE, apr 2009, pp. 1--4. [Online].
  Available: \url{http://ieeexplore.ieee.org/document/5347957/
  http://ieeexplore.ieee.org/lpdocs/epic03/wrapper.htm?arnumber=5347957}
\BIBentrySTDinterwordspacing

\bibitem{Machowski2008}
J.~Machowski, J.~W. Bialek, and J.~R. Bumby, \emph{{Power System Dynamics -
  Stability and Control}}, 2nd~ed.\hskip 1em plus 0.5em minus 0.4em\relax John
  Wiley {\&} Sons, Ltd, 2008.

\bibitem{Liu2014UseThis}
\BIBentryALTinterwordspacing
S.~Liu, Z.~Xu, W.~Hua, G.~Tang, and Y.~Xue, ``{Electromechanical Transient
  Modeling of Modular Multilevel Converter Based Multi-Terminal HVDC
  Systems},'' \emph{IEEE Transactions on Power Systems}, vol.~29, no.~1, pp.
  72--83, jan 2014. [Online]. Available:
  \url{http://ieeexplore.ieee.org/document/6588378/}
\BIBentrySTDinterwordspacing

\bibitem{paulo2012}
P.~Chainho, A.~van~der Meer, M.~Gibescu, R.~Hendriks, and M.~van~der Meijden,
  ``{General modeling of multi-terminal VSC-HVDC systems for transient
  stability studies},'' in \emph{Sixth IEEE young researchers symposium in
  electrical power engineering. Challenges in sustainable electrical energy},
  2012, pp. 1--6.

\bibitem{Shewarega2014}
F.~Shewarega and I.~Erlich, ``{Simplified Modeling of VSC-HVDC in Power Systems
  Stability Studies},'' in \emph{The19th World Congress The International
  Federation of Automatic Control}, 2014.

\bibitem{olofsson1996}
M.~Olofsson, ``{Optimal Operation of the Swedish Railway Electrical System - An
  application of Optimal Power Flow},'' Ph.D. dissertation, Royal Institute of
  Technology, 1996.

\bibitem{Simpow2010}
Stri/Solvina, \emph{{Simpow Manual 11.0}}, 2010.

\bibitem{IEC61000430}
IEC, ``{Electromagnetic compatibility (EMC) - Part 4-30: Testing and
  measurement techniques - Power quality measurement methods},'' 2015.

\bibitem{Fridman2006}
E.~Fridman, ``{Impedanser f{\"{o}}r/ Impedances KTL och 132 kV, 30 kV och 15 kV
  ML},'' Trafikverket, Tech. Rep., 2006.

\end{thebibliography}
}







\end{document}